# Hiding Secret Information in Movie Clip: A Steganographic Approach

*G. Sahoo*
Department of Information Technology., B. I. T. Mesra, Ranchi, Jharkhand, India
*Rajesh Kumar Tiwari*
Computer Sc. & Engg. Dept., R. V. S. College of Engg. & Tech., Jamshedpur, Jharkhand, India

*Establishing hidden communication is an important subject of discussion that has gained increasing importance nowadays with the development of the internet. One of the key methods for establishing hidden communication is steganography. Modern day steganography mainly deals with hiding information within files like image, text, html, binary files etc. These file contains small irrelevant information that can be substituted for small secret data. To store a high capacity secret data these carrier files are not very supportive. To overcome the problem of storing the high capacity secret data with the utmost security fence, we have proposed a novel methodology for concealing a voluminous data with high levels of security wall by using movie clip as a carrier file.*

***Index Terms****: Steganography, Frame, Stego-Frame, Stego-key, and Carrier.*

## 1. INTRODUCTION

The growing use of internet needs to store, send or receive personal information. The common approach is to transfer the data in different form so that the resultant data can be understood only by those who can get it back into its original form. This technique is known as encryption. A major draw back of this method is that the existence of data is not hidden. Giving enough time the unreadable encrypted data can be obtained in its original form.

As a part of solution to this kind of problems, the concept of steganography has considerably given much attention in recent years. The word steganography is a Greek word giving a meaning to it as 'writing in hiding'. The main purpose of steganography is to hide data in a cover media so that other will not notice it[10]. The characteristics of cover media depends on the amount of data that can be hidden, the perceptibility of the message and its robustness [4, 5, 6, 8, 10].

Fields like publishing and broadcasting fields also require an alternative solution for hiding information. Unauthorized copying becomes a hot issue in the area like music, film, print and software. To overcome such type of problems some invisible information can be embedded in the digital media in such a way that no one easily extract it [3,11, 12, 14].

## 2. Related Works

The most suitable cover media for Steganography is an image. As a result, numerous methods have been used. The main reason behind it is the large redundancy and the possibility of hiding information in the image without attracting attention to human visual system. In this respect, a number of techniques have been developed [13, 15] using features like

- Substitution
- Masking & Filtering
- Transform Technique

The method of substitution generally does not increase the size of the file. Depending on the size of the hidden message, it can eventually cause a noticeable change from the unmodified version of the file [1, 2, 4, 6]. In this regard, the Least Significant Bit (LSB) insertion technique is an approach for embedding information in an image file. Where, every least significant bit of some or all of the bytes inside an image is changed to a bit of the secret message.

In case of context masking and filtering techniques, starts with the analysis of the image and then, we find the significant areas, where the hidden message will be more integrated to cover the image and finally we embed the data in that particular area. Therefore, while in the case of LSB techniques all the least significant bits are changed, in masking and filtering techniques we just say that the changes can take place only in selected areas.

In addition to the above techniques for message hiding, transform techniques has also played some important role in embedding the message by modulating coefficients in a transform domain. As an example, Discrete Cosine Transform works by using quantization on the least important parts of the Image in respect to the human visual capabilities. Marvel [6] has proposed a spread spectrum image steganographic technique for image carrier and able to conceal 5 kB of secret data in 512 x 512 image size. Julio C. Hernandez-Castro [7] has given a concept for steganography in games where the less robust technique saves only few parts of total carrier files. Raja [8] concluded that three secret bits can be embedded in one pixel. Amin [10] has given a secured information hiding system which can only embed 60 kB message in 800 x 600 pixels image.

In conclusion, we can say that all of the above methods and carriers are having low embedding characteristics. To conceal a voluminous secret data we require a large capacity carrier files and advanced methodology for the purpose of embedding. In the proposed methodology, we are considering a movie clip as a carrier and the new methodology that works on the concept of replacement of entire non- sensitive pixel and the substitution of some part of the sensitive pixel with the secret data. The paper is organized as follows. The proposed method is discussed in the following section. Section 4 describes the process of data retrieval. Followed by conclusion in section 5.

### 3. PROPOSED METHOD

In multimedia systems, a variety of information sources are delivered synchronously or asynchronously via more new invented hardware devices. The rapid changes in computer hardware world with high speed processors, high capacity primary and secondary storage devices give the plenty of chance for computing the large volume of data. The different phases are as follow for the embedding secret data as shown in Fig. 1.

**Phase-I (Secret Data Analysis & Clip Slicing)**

The secret data, which the sender wishes to keep confidential, is the original message or data that is fed into the algorithm considered as input. Here, we start our process by analyzing the size of the secret data. Based on the size of secret data we move to the selection of the size of the movie clip. Generally we consider a larger clip for the embedding purpose. Movie segmentation is a very important step for the efficient secret data embedding. This is due to fact that a digitized movie video data can be several gigabytes in size. A movie can be divided into video clips and these clips can be partitioned into scenes. A scene is a common event or locale which contains a sequential collection of shots. A shot is a second basic unit of video clip, and the last but not the least the contiguous frame is the primary unit of the movie clip. A movie clip is temporal sequence of two dimensional samples of the visual field with each sample being a frame is referred to as a frame of movie. The samples are acquired at regular intervals of time which is currently 30 frames per second for most of the standards movie matching devices (Fig. 2.)

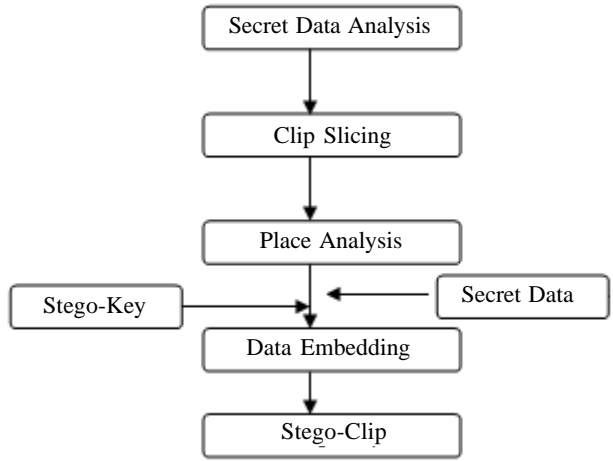

**Figure 1: Basic Model**

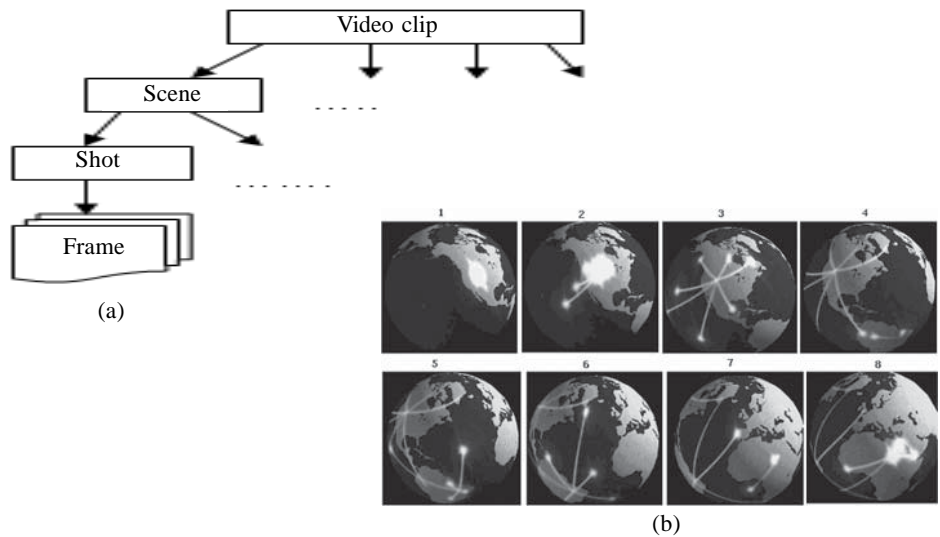

**Figyre 2: (a) Hierarchy of Video Clip. (b) Frames of a Shot**

**Phase-II (Place Analysis)**

The primary unit of the movie clip is the ultimate region where we embed the data. We select the suitable place in the frame for the embedding process. Practical results show that moving part of any clips is more interactive, attractive and effective for the human visual system (HVS) whereas, the static portions of the clip are less attractive. We concentrate in the static portions of the two or more consecutive frames. The static portions of the frames give the freedom to embed more secret data with very minor changes, and, for dynamic portion of frames, we are using the least significant bits technique. The static and dynamic portion of the frame may be obtained by the following methods.

(a) *Pixel Level Analysis Technique:* In this technique, gray-scale values of pixels at corresponding locations of two or more consecutive frames are analyzed. The difference of the pixels intensities gives the exact information whether the portion are static (similar) or not. If the resultant is approximately zero, it shows that the portions in different frames are static in nature and we can embed the large data there.

(b) *Likelihood Analysis Technique:* The other option for the comparison of static and dynamic portion of the consecutive frames may be used for the embedding process. In this technique, the frames are divided into small blocks. The blocks of two consecutive frames are compared based on some statistical characteristics of their intensity values and based on this, we find the suitable place for the embedding the large secret data.

(c) Color histogram Technique: In this approach, a frame is analyzed by dividing the color space into discrete colors called color basket and counting the number of pixels that fall into each color basket. For this a separate histogram may be made for each primary color.

Based on the above techniques discussed we can find the static and dynamic portion of the frames and the resultant location is stored in the static and dynamic buffer respectively (Fig. 3).

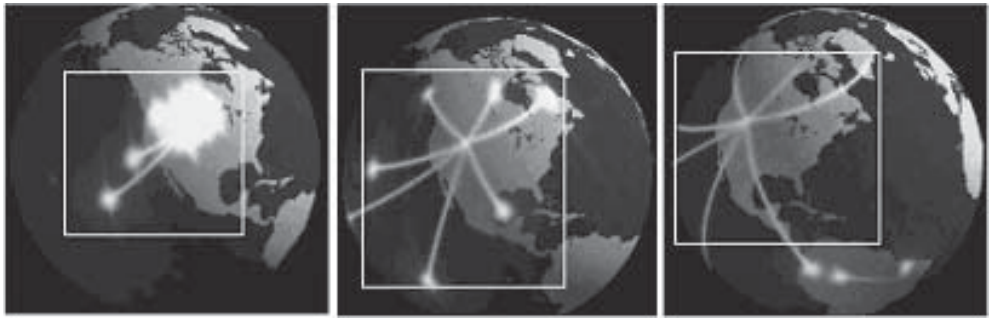

Figure 3: Dynamics Portion of the Frame is Marked with Yellow Rectangle

**Phase-III (Stego-Key & Data Embedding)**

The stego-key is known as password, which is a protection fence for this methodology. The analyzed secret data is the second input for this methodology. The embedding process may be divided into two processes, the static embedding technique takes care for the static buffer value and the second technique takes care for the dynamic buffer value.

**Hiding Secret Information in Movie Clip: A Steganographic Approach** 91

(a) *Static Portion Embedding Process:* Our main aim here is to conceal the entire one character in place of one byte of pixel value with fewer amounts of changes. Here one pixel can be used to store three characters, whereas by LSB technique one need a minimum of nine pixels to store these three characters as three pixels are required to store a single character. Before the embedding process, we check the total size of the static buffer and the secret data length. For introducing an extra layer of protection fence we may use mathematical formula for embedding the secret data into the static portion of the frames. For illustration purpose we consider the mathematical formula as

$$X_{ij} = i + (j - 1) * d \qquad \ldots \text{(i)}$$

such that if the first character of secret data $j = 1$, the initial location $i = 5$, and the distance between two embedding pixels $d = 3$ then the first secret character will be stored at $X_{ij} = 5^{th}$ byte place, second characters will be stored at $8^{th}$ byte place and so on. This process gives opportunity to embed the three characters per pixel which boost up the high embedding capacity. The required pseudo code is given below.

```
/* Pseudo Code for Static Portion Embedding Method */
Static Embedding Method
{
1DArray Cover_Frame= Load ("Cover Frame", &length);
1D Array Stego_Frame [ ];
1D Array Secret_data [ ];
1D Array ASCII_data [ ];
Integer R, G, B;
Integer I, Pixel_Place;
INPUT Secret _data
/* Convert Secret data into corresponding ASCII value */
Bit_Conversion (ASCII_data [ ], Secret_Data [ ]);
/* Secret data Storing * /
If Length (Secret_data [ ]) > Length (Cover_Frame [ ]) Then
    Print "Secret Data size is more … "
Else
Pixel_Place=0
   For I =0 to Length (Secret_data [ ])
   GetColorValue (&R, &G, &B, Cover_Frame [Pixel_Place]);
   R= ASCII_data [I]; G= ASCII_data [I + 1];
   B= ASCII_data [I +2];
   SetColorValue (Stego_Frame [Pixel_Place], &R, &G, &B);
   Pixel_Place= Pixel_Place+1;
   I=I+2; End For End If;
SaveFrame ("Stego_Frame");
}
```

(b) *Dynamic Portion Embedding Process:* The majority of the secret data embedding may be done by the static portion embedding process. If there are still some secret data remains, it can be concealed by using the most useful most significant bits method. This will lead a very minute difference on the interactive and sensitive portion of the frame. For highlighting the above fact, we use here two different techniques for embedding the secret data for static and dynamic portion of the frames. We also use two different types of secret data and the stego-key for static and dynamic portions and may mention double role of a single carrier file (Fig. 4). The required pseudo code is given below.

```
/* Pseudo Code for Dynamic Portion Embedding Method */
Dynamic Embedding Method
{
1DArray Cover_Frame= Load ("Cover Frame", &length);
1D Array Stego_Frame [ ];
1D Array Secret_data [ ];
1D Array ASCII_data [ ];
Integer I, J=0, k;
Input Secret_Data;
/*Secret Data bit Conversion by using predefined or library functions */
   Bit_Conversion (ASCII_data [ ], Secret_Data [ ]);
For I =1 to Length (ASCII_data [ ])
   If ASCII_data [I] <> MOD (Cover_Frame[k],2) then
   If ASCII_data [I] =0 Then
      Cover_Frame[k] = Asc (Cover_Frame[k])-1;
   Else
      Cover_Frame[k] = Asc (Cover_Frame[k])+1;
   End If; End If; k=k+1; End For
SaveFrame ("Stego_Frame");
}
```

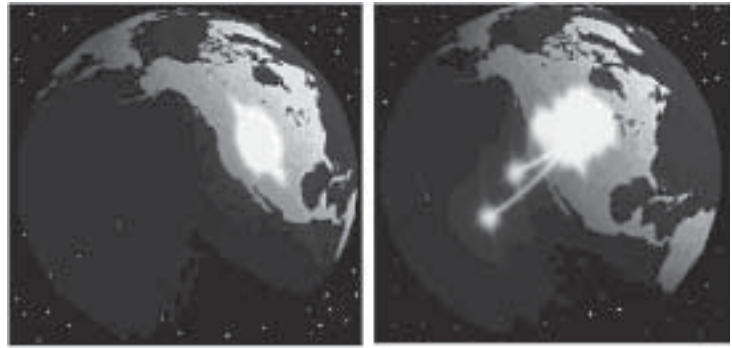

**Figure 4: Stego-clip**

## 4. DATA EXTRACTION & DISCUSSION

The embedded stego frame can be retrieved back by using an extraction algorithm. Generally, the extraction is the reverse process of embedding the data in to the cover frame. The stego frame is the primary input for data extraction algorithm. The retrieval process takes the inputs of initial pixel place, stego key and the mathematical formula by which the secret data has been distributed. The next pixel location is obtained by the mathematical formula being generated and the corresponding pixel ASCII value is converted into the characters. Finally, with the aid of a grammar and dictionary the transmitting data is retrieved and fabricated.

The three different aspects considered in any information hiding system are capacity, security, and robustness. Capacity refers to the amount of information that can be hidden in the cover medium; security refers to an eavesdropper's inability to detect hidden information whereas the robustness is referred to the amount of modification that can be made such that the stego medium can withstand before an adversary who can destroy the hidden information. In this proposed system, as we are using the unbounded carrier of size that may be in giga-bytes, we can embed as much data as we want and also maintain three level of security one from stego-key, second from mathematically distributed secret data and the last from the portioning of frame into static and dynamic parts. Moreover, this methodology creates a robust stego movie clip.

## 5. CONCLUSION

Adopting a new methodology, the suitability of steganography as a tool to conceal highly sensitive high capacity secret data has been discussed. The new methodology suggests that an unbound carrier like movie clip may be an excellent source for concealing the high capacity data. The two embedding techniques provide the opportunity to use the carrier for dual role. The stego video clip can be transmitted any where across the world in a complete secure form. Industries like music, film, publishing and organization like ministry and military will definitely be highly benefited by the using such techniques.